\def\Xint#1{\mathchoice
   {\XXint\displaystyle\textstyle{#1}}%
   {\XXint\textstyle\scriptstyle{#1}}%
   {\XXint\scriptstyle\scriptscriptstyle{#1}}%
   {\XXint\scriptscriptstyle\scriptscriptstyle{#1}}%
   \!\int}
\def\XXint#1#2#3{{\setbox0=\hbox{$#1{#2#3}{\int}$}
     \vcenter{\hbox{$#2#3$}}\kern-.5\wd0}}

\def\dashint{\Xint-}

\documentclass[aps,prb,twocolumn,showpacs,floatfix,eqsecnum]{revtex4}

\pdfoutput=1

\usepackage{times}
\usepackage{amsfonts}
\usepackage{amssymb}
\usepackage{amsmath}
\usepackage{graphicx}
\usepackage{bm}
\usepackage{verbatim}

\usepackage{hyperref}

\usepackage{bm} 
\usepackage{color}
\usepackage{stackrel}
\usepackage{accents}

\usepackage{latexsym}

\newcommand{\bsub}{\begin{subequations}}
\newcommand{\esub}{\end{subequations}}

\newcommand \bea {\begin{eqnarray} }
\newcommand \eea {\end{eqnarray}}
 
\newcommand{\beg}{\begin{equation}}
\newcommand{\en}{\end{equation}}
\newcommand{\bp}{\mathbf p}
\newcommand{\bq}{\mathbf q}
\newcommand{\bk}{\mathbf k}

\newcommand \bel  {\begin{align}}
\newcommand \enl  {\end{align}}

\newcommand{\eps}{\varepsilon}

\newcommand{\up}{\uparrow}
\newcommand{\dn}{\downarrow}
\newcommand{\dg}{^\dagger}

\renewcommand{\Re}{\mathrm{Re}}

\newcommand{\pmat}{\begin{pmatrix}}
\newcommand{\epmat}{\end{pmatrix}}

\def\8{\infty}

\def\undertext#1{\vtop{\hbox{#1}\kern 1pt \hrule}}

\def\be{\begin{equation}}
\def\ee{\end{equation}}
\def\bea{\begin{eqnarray} & &}
\def\eea{\end{eqnarray}}


\begin{document}

\title{Non-adiabatic dynamics in $d+id$-wave fermionic superfluids}

\author{Ammar A. Kirmani and Maxim Dzero}
\affiliation{Department of Physics, Kent State University, Kent, OH 44242, USA}
\begin{abstract} 
We consider a problem of non-adiabatic dynamics of a 2D fermionic system with $d+id$-wave symmetry of paring amplitude. Under the  mean-field approximation, we determine the asymptotic behavior of the pairing amplitude following a sudden change of coupling strength. We also study an extended $d+id$ pairing system for which the long-time asymptotic states of the pairing amplitude in the collisionless regime can be determined exactly. By using numerical methods, we have identified three non-equilibrium steady states described by different long-time asymptotes of the pairing amplitude for both the non-integrable and the integrable versions of $d+id$-wave models. We found that despite of its lack of integrability, long-time dynamics resulting from pairing quenches in the non-integrable $d+id$ model are essentially similar to the ones found for its exactly-integrable extended $d+id$ model. We also obtain the long-time phase diagram of the extended $d+id$ model through the Lax construction that exploits underlying integrability showing that the dynamic phases obtained by numerics are consistent with the dynamics of the exactly integrable approach. Both models describe a topological fermionic system with a topologically non-trivial BCS phase appearing at weak coupling strength. We show that the presence of oscillating order parameter region in the chiral $d+id$ pairing dynamics differs from the d-wave ($d_{x^2-y^2}$), which may be used to probe pairing symmetries of chiral superconductors.  
\end{abstract}


\maketitle

\section{Introduction}
Exactly solvable models of many-body quantum systems have always been a powerful tool for developing important ideas about the nature and the microscopic structure of physical phenomena especially when inter-particle interactions are strong. Integrability puts stringent constraints on the conditions under which models are formulated (reduced dimensionality, purely local interactions etc.) rendering physical systems often to be in extreme physical situations. Nevertheless, the concepts developed by using exact solutions make it possible to gain deeper insight into complex physical phenomena and are fruitfully applied to provide interpretation of underlying physical ideas. 

Among the exactly-integrable Hamiltonians, the Gaudin magnets \cite{Gaudin_book,Sklyanin1989,Sklyanin1995,Emil2005a} represent a special type of integrable many-body systems formulated in terms of the spin Hamiltonians. It is well known that within the mean-field approximation, the Hamiltonian for the celebrated BCS model can be formulated as Gaudin spin Hamiltonian and, therefore, is exactly integrable.\cite{Emil2005a,Emil2005b,Emil2006} This fact turned out to be especially useful for solving the problem of non-adiabatic pairing in fermionic superfluids (for review see Ref. [\onlinecite{Big-Quench-Review2015}] and references there in).

Since the discovery of the exact solution for the non-adiabatic pairing problem, there has been a lot of theoretical studies addressing various  related aspects of the problem. For example, steady states for different types of pairing symmetries such as the chiral $p$-wave,\cite{Yuzbashyan2013,Foster2014} the $d$-wave,\cite{Capone2015} effects of the various integrability breaking perturbations on dynamics phase diagram \cite{BarankovGij} as well as dynamics in two-dimensional spin-orbit coupled fermionic superfluids in external Zeeman field \cite{HanPu2014,Dzero-2dso-2015} have been discussed. Perhaps the most remarkable results of many of these studies is that breaking of integrability does not lead to the substantial deviations from the results found for the integrable model.\cite{Dzero-2dso-2015} 

Although experimental observation(s) of non-adiabatic pairing phenomena in degenerate atomic condensates is still lacking, 
there have been significant advances in realization of non-adiabatic pairing regime in superconductors by employing the technique of pump-probe spectroscopy. While it has been shown experimentally that superconducting state can be photo-induced \cite{Cav2011Science,Shimano2012,Shimano2013,CavNature2014,Shimano2014Science,CavPRB2015}, it was not until recently shown that when pump pulses are used in combination with angle resolved photo-emission spectroscopy (ARPES), real-time evolution of quasiparticle modes following a pump pulse can be tracked in different momentum sectors of $d$-wave cuprates superconductors, revealing highly non-thermal character of the associated spectral weights even for the steady state asymptote of order parameter. \cite{RefE,RefF}\\
The fact that there have been several proposals on realization of unconventional $s+is$- and $d+id$-wave pairings in multiband superconductors,\cite{sisEremin2017} motivates us to look into the signatures of such an unconventional pairing in non-adiabatic regime. Pairing problem with $d_{x^2-y^2}+id_{xy}$ ($d+id$) pairing symmetry of the order parameter has received attention recently in the context of possible topologically nontrivial superconducting states in undoped bilayer silicene and to explain Broken Time Reversal Symmetry (BTRS) in  YBa2Cu3Ox superconductors. Generally, the $d+id$ pairing has a number of highly unusual physical properties such as quantized boundary current, spontaneous magnetization as well as quantized spin and thermal Hall conductances. \cite{LaughlinDID,HorovitzDID,LiuDID} Furthermore, It has also been shown in \onlinecite{HorovitzDID}, that $d+id$ superconductor leads to spontaneous magnetization which is temperature independent for the weak BTRS in accordance with experiments \cite{NatureLondon} . Transitions of YBa2Cu3O7-x films, from pure $d_{x^2-y^2}$ to $d+id$ was also proposed based on experimental observations \cite{Dagan2001}. Change of d-wave paring to $d_{x^2-y^2}+id_{xy}$ is also attributed to plateaus observed in the field profile thermal conductivity measurements in Bi2Sr2CaCu2O8 \cite{Krishana83}. Moreover, chiral superconductivity from repulsive interactions in doped graphene has also been proposed in the context of $d+id$ pairing.\cite{Nandi2012}\\\\
In this paper we consider a problem of non-adiabatic dynamics of systems with $d+id$ symmetry of the order parameter. Within the mean-field approximation, the $d+id$-wave model Hamiltonian can be written as a spin Hamiltonian, however it does not belong to the class of Gaudin magnets and hence it is not integrable. With inclusion of an extra term, the model becomes integrable. Thus, we can establish similarities and differences between the corresponding asymptotic phase diagrams and the related observables for the two models. 

{Our paper is organized as follows. In Section II we introduce the Hamiltonian for the $d+id$ model, describe its mean-field and topological properties. In Section III, we present dynamics of the non-integrable $d+id$ and the extended $d+id$ in weakly couples regime. In section IV, we establish the integrability of the extended $d+id$ model and using the Lax vector construction, obtain the corresponding exact asymptotic state phase diagram. Section V is devoted to the discussion of our results and conclusions.} 
\section{Model Hamiltonian and Ground State}
\begin{figure}[ht]
\centering
\includegraphics[width=9cm]{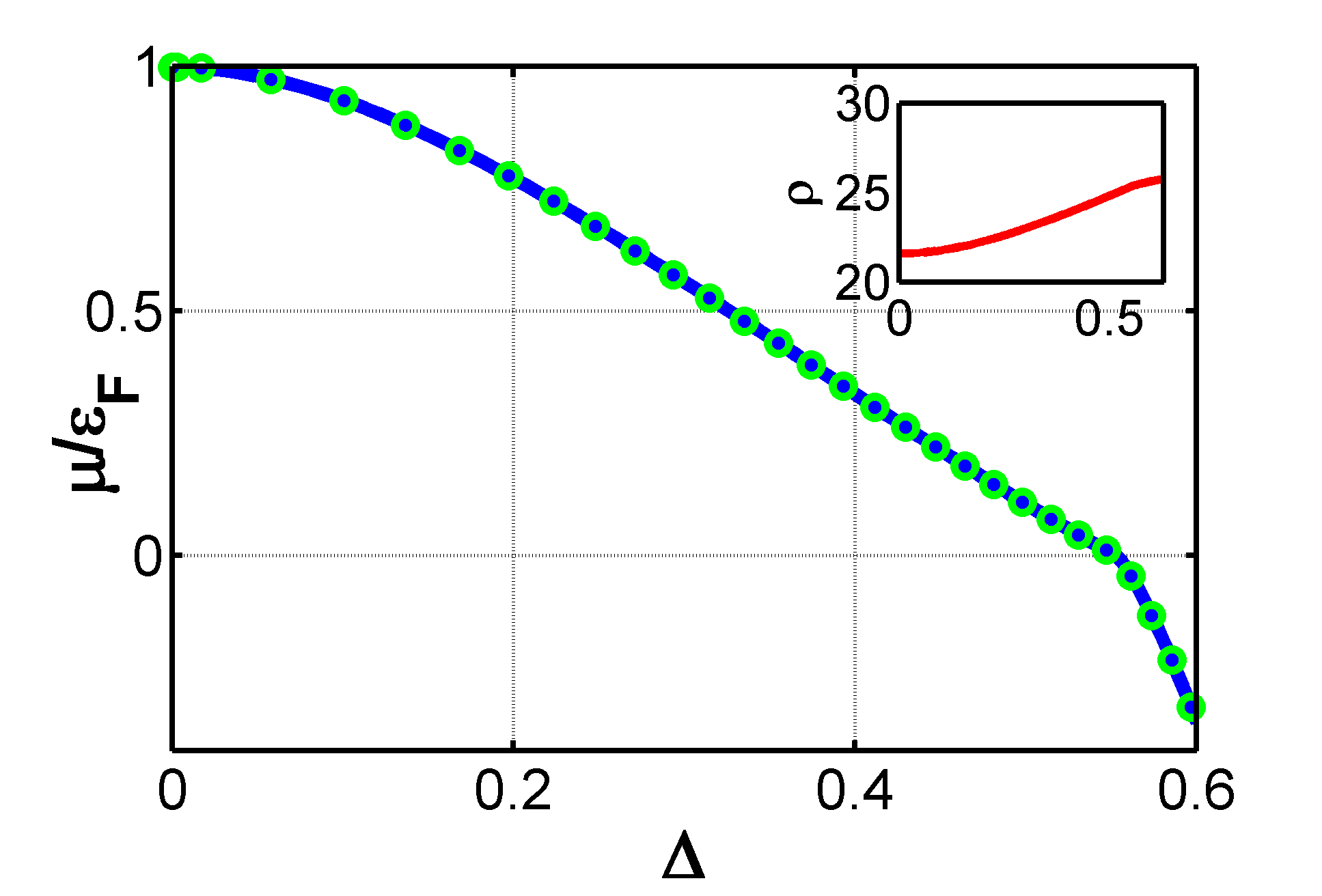}
\caption{Mean-field ground state values of chemical potential $\mu$ and order parameter $|\Delta|$ for $d+id$ and extended $d+id$ model. Solid blue curve gives $\mu$ vs $\Delta$ for $d+id$ and green circles gives $\mu'=\mu/(1+\rho)$ vs $\Delta'=\Delta/(1+\rho)$ for extended $d+id$ (inset $\rho$ vs $\Delta'$). To solve system of equations (\ref{Self},\ref{zeta} and \ref{ParticleNumber}), we have implemented the ultraviolet cutoff $\Lambda=16\varepsilon_F$ ($\varepsilon_F$ is the Fermi energy).}
\label{Fig:SolveSelf}	
\end{figure}

An exact ground solution of the s-wave Hamiltonian was obtained through the series of work pioneering in the field. \cite{Richardson1,Richardson2,GaudinExact} Furthermore, it has been shown that there are Hamiltonians beyond the s-wave case which can be solved exactly in any dimensions. \cite{RefA,RefB} One of such cases, is the $p+ip$-wave pairing model with a non-trivial ground state and topological properties, when solved exactly, is in agreement with the mean-field analysis.\cite{RefC,RefD} Next in line of the $p+ip$ Hamiltonian, the chiral $d+id$ is non-integrable and do not admit an exact solution whereas integrability of the extended-$d+id$ is established and shown in earlier work that the exact ground state analysis obtained through the Bethe ansatz agrees with the mean-field solution. \cite{Links2013}\\ 
Non-integrable case of the $d$-wave pairing order parameter dynamics has been presented in \onlinecite{Capone2015}, where it has been shown that out-of-equilibrium spectral weight along the nodal lines obtained by the mean-field calculations qualitatively behaves in a similar manner as reported by the experiments. \cite{RefE,RefF} Integrability has been exploited in earlier work including the s-wave  and the 2-D $p+ip$-wave Hamiltonians to compute non-adiabatic order parameter dynamics following a quench of interaction constant. It was shown, when the ground state has non-trivial topology, integrability even allows to compute non-equilibrium topological invariant  \cite{Big-Quench-Review2015}\cite{Yuzbashyan2013}. Nevertheless, link between integrability and out-of-equilibrium dynamics is still a subject of further studies. In this Section, we present our results for the mean-field ground state of the non-integrable $d+id$ and the integrable extended-$d+id$ model.\\
We introduce the singlet BCS Hamiltonian with the $d+id$-wave pairing symmetry \cite{ReadGreen2000} \cite{Sato2010}
\beg\label{Eq1}
\begin{split}
\hat{H}&=\sum\limits_{\bk\sigma}\eps_\bk\hat{c}_{\bk\sigma}\dg\hat{c}_{\bk\sigma}-\frac{G}{4\nu_F^{-1}}\sum\limits_{\bk,\bq}k_{+}^2q_{-}^2\hat{c}_{\bk\up}\dg\hat{c}_{-\bk\dn}\dg\hat{c}_{-\bq\dn}\hat{c}_{\bq\up},
\end{split}
\en
where $\hat{c}_{\bk\sigma}$ and  $\hat{c}\dg_{\bk \sigma}$ denote fermionic annihilation and creation operators, 
$\eps_\bk={\bk^2}/{2}$ is the single particle dispersion, $k_\pm=k_x\pm ik_y$, $G$ is a dimensionless coupling constant and $\nu_F$ is the density of states at the Fermi level. We set the single particle mass $m=1$. Given the fact that the second term in the Hamiltonian can be factorized, it is convenient to write it in terms of the operators 
\beg\label{Pseudospins}
\begin{split}
&\hat{S}_\bk^+=\hat{c}_{\bk\up}\dg\hat{c}_{-\bk\dn}\dg, ~\hat{S}_\bk^{-}=\hat{c}_{-\bk\dn}\hat{c}_{\bk\up},\\ 
&\hat{S}_\bk^z=\frac{1}{2}\left(\hat{c}_{\bk\up}\dg\hat{c}_{\bk\up}+\hat{c}_{-\bk\dn}\dg\hat{c}_{-\bk\dn}-1\right).
\end{split}
\en
These are the familiar Anderson pseudospin operators \cite{Anderson1958} which satisfy the angular momentum commutation relations $[S^a_\bk,S^b_\bq]=i \epsilon^{abc}\delta_{\bk\bq}S^c_\bk$ provided the momentum summation is restricted 
to the range 
\beg\label{MomentumRestrict}
\bk=\{k^x \in \Re, k^y\geq 0\}.
\en 
This model only considers sub-space of paired fermions related by time reversal symmetry and neglects pair-breaking processes, giving us:
\beg\label{ASH}
\hat{H}_{d+id}=2\sum\limits_{\bk}\eps_{\bk}\hat{S}_\bk^z-\frac{G}{\nu_F^{-1}}\sum\limits_{\bk,\bq}\eps_{\bk}\eps_{\bq} \hat{S}_\bk^{+} 
\hat{S}_\bq^{-}.
\en
Where we have also eliminated the momentum phase prefactors $(k_x\pm ik_y)^2=|\bk|^2e^{\pm2i\phi_\bk}$ by performing unitary transformation for the pseudospin operators and absorbed constant into interaction parameter. Hamiltonian in Eq. (\ref{ASH}) is non-integrable and an exact solution does not exist. We add an extra term in Eq. (\ref{ASH}) proportional to density-density interaction to arrive at an extended-$d+id$ model. \cite{Links2013}
\beg\label{ASHE}
\hat{H}_{ex}=2\sum\limits_{\bk}\eps_{\bk}\hat{S}_\bk^z-\frac{G}{\nu_F^{-1}}\sum\limits_{\bk,\bq}\eps_{\bk}\eps_{\bq} \hat{S}_\bk^{+} 
\hat{S}_\bq^{-}-\frac{G'}{\nu_F^{-1}}\sum\limits_{\bk,\bq}\varepsilon_\bk\varepsilon_\bq S_\bk^z S_\bq^z
\en
The exact solution of (\ref{ASHE}) was obtained in \onlinecite{Links2013}, where it was also shown that the system is integrable in case when $G'=G$. Moreover, it has also been shown in the same work that the exact solution coincides with the mean-field case. To obtain the ground state in the mean-field approximation, the pseudospin operators are replaced with their expectation values $\hat{S}_\bk^a\to\langle \hat{S}_\bk^a\rangle={S}_\bk^a$. As a result, the spin Hamiltonian (\ref{ASHE}) 
becomes a classical Hamiltonian of the form
\beg\label{Hmf}
H_{ex}=\sum\limits_{\bk}{\vec B}_\bk\cdot{\vec S}_\bk, ~{\vec B}_\bk=2(-\Delta_x \eps_\bk,-\Delta_y\eps_\bk,\xi_\bk)
\en
where $\xi_\bk=\eps_\bk(1+\rho)-\mu$ and $\Delta_{x,y}$ are the components of the complex pairing field 
\beg\label{Self}
\Delta^\dagger=\Delta_x+i\Delta_y=(G/\nu_F^{-1})\sum_\bk\eps_\bk S_{\bk}^+. 
\en
along with the parameter
\beg\label{zeta}
\rho=-\frac{G'}{\nu_F^{-1}}\sum\limits_{\bp}\varepsilon_\bp S_\bp^z.
\en 
We set, for the non-integrable $d+id$ case $G'=0$ ($\rho=0$) whereas $G'=G$ for the integrable version. Time evolution of the pseudospin components along with the pairing field $\Delta^{+}$ is governed by the classical equations of motion which are obtained by evaluating the Poisson brackets of $S_\bk^a$ with the Hamiltonian:
\beg\label{EqMot}
\begin{split}
	\dot S_{\bk}^x= -2\eps_{\bk} \Delta_{y}(t)S_{\bk}^z(t)-2\xi_{\bk}S_{\bk}^y(t), \\
	\dot S_{\bk}^y= 2\xi_{\bk}S_{\bk}^x(t)+2\eps_{\bk} \Delta_{x}(t)S_{\bk}^z(t),\\
	\dot S_{\bk}^z=-\eps_{\bk}2\Delta_x(t)S^y_{\bk}(t)+2\eps_{\bk}\Delta_y S^x_k(t).\\
\end{split}
\en
Above equations can be summarily written as $\dot{\vec{S}}_\bk(t)=\vec{B}_\bk(t) \times \vec{S}_\bk(t)$. In the ground state, each pseudospin is aligned so that the time derivatives in Eq. (\ref{EqMot}) are identically zero. For simplicity we assume that in the ground state $\Delta=\Delta_x$, it follows
\beg\label{Equi}
S_\bk^x=\frac{\eps_{k}\Delta }{2\sqrt{\xi_{k}^2 + |\eps_{k}\Delta|^2}}, ~
S_\bk^z=-\frac{\xi_{k}}{2\sqrt{\xi_{k}^2 + |\eps_{k}\Delta|^2}}
\en
and $S_\bk^y=0$. In addition to the self-consistency equation(s) for the pairing field, we also need to consider the particle number equation which fixes the value of chemical potential:
\beg\label{ParticleNumber}
n=\sum_{\bk}\left(1-\frac{\xi_{k}}{\sqrt{\xi_{k}^2 + |\eps_{k}\Delta|^2}} \right).
\en
For ground state,  $\rho$ in Eq. (\ref{zeta}) renormalizes the chemical potential and the order parameter of our system: $\mu\to \mu/(1+\rho)$ and $\Delta\to \Delta/(1+\rho)$ ($\rho=0$ for the non-integrable $d+id$). We solve Eqs. (\ref{Self},\ref{zeta},\ref{ParticleNumber}) numerically and show results in Fig. \ref{Fig:SolveSelf}. 
As it turns out, just as in the case of the chiral $p+ip$ case, \cite{ReadGreen2000,Yuzbashyan2013,Foster2014,RefC,RefD}
the point $\mu=0$ is a special one: it marks the transition between the two topologically distinct states. From Fig. \ref{Fig:SolveSelf}, it is evident that the ground state pseudospins for the non-integrable and the integrable case are identical. The topological invariant or winding of pseudospin configuration is independent of value of $G'$ and in 2D is given by \cite{Yuzbashyan2013}
\beg\label{TI}
Q=8\pi \eps_{abc}\int \frac{d^2\bk}{(2\pi)^2}\frac{1}{k} \langle s^a_\bk\rangle \partial_k\langle s^b_\bk \rangle \partial_{\phi_k} \langle s^c_\bk\rangle
\en 
We can proceed with the substitution, $\beta_\bk=-\xi_{\bk}/E_\bk$ and obtain form of equilibrium pseudospins from Eq. \ref{Equi}: 
\beg\label{TI_PS}
\begin{split}
S_\bk^z=\frac{ \beta_\bk}{2}, \quad
S_\bk^\pm=\frac{1}{2}\sqrt{1-\beta_\bk^2}e^{\mp 2i\phi_\bk},\\
\end{split}
\en
With the help of expressions (\ref{TI_PS}), the integrand in Eq. (\ref{TI})
reduces to total derivative, giving us 
\beg
Q=\left\{
\begin{matrix} 
2, \quad \mu>0 , \\
0, \quad  \mu<0
\end{matrix}\right.
\en
in the ground state. Non-zero winding number signals the presence of the Majorana edge states at the system's boundaries, so $Q=2$ implies that the chiral $d+id$-wave system (weakly coupled) supports two of these edge modes - one per each spin projection. This give a quantized boundary current of $I_B=2e\Delta/h$.\cite{LaughlinDID} Chiral $d+id$ pairing is important in context of superconductors with other usual characteristics that cannot be explained by the d-wave pairing. Thus we focus on the weakly coupled (BCS) dynamics where $\mu \approx 1 \eps_F$.
\section{Out-of-equilibrium dynamics: phase diagram}

\begin{figure}[h]
	\centering
	\includegraphics[width=9cm]{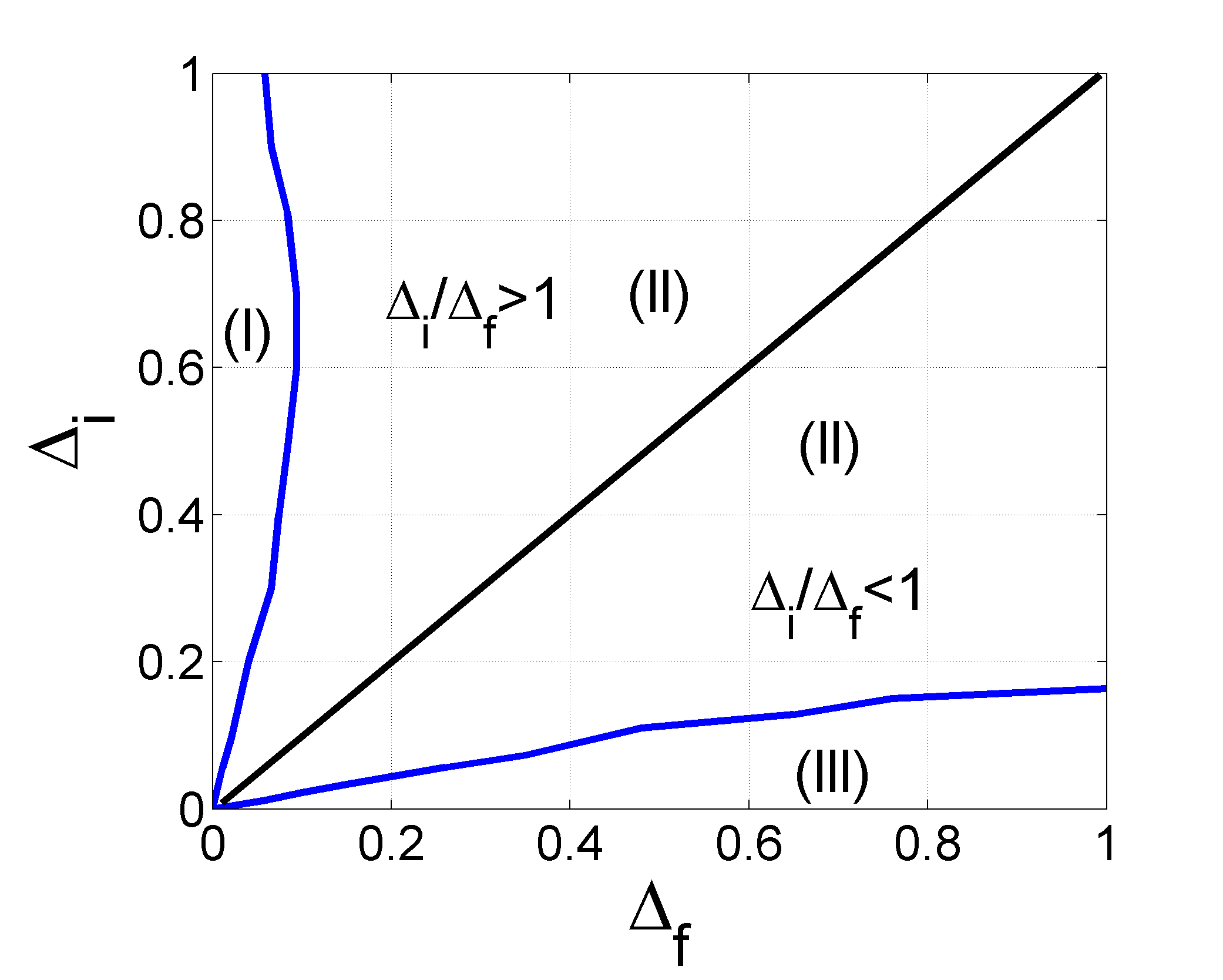}
	\caption{Non-integrable $d+id$ ($G'=0$) phase diagram obtained from quenches showing the three dynamic regions where $|\Delta(t)|$ can be classified at long times. Region I is characterized by vanishing $|\Delta(t)|$, Region II gives non zero  asymptote for $|\Delta(t)|=\Delta_{\infty}\neq 0$ and finally Region III gives persistent oscillations of order parameter. Computations have been performed with $\Lambda_E=4\eps_F$.} 
	\label{fig:didpd}	
\end{figure}

\begin{figure}[h]
	\centering
	\includegraphics[width=7cm]{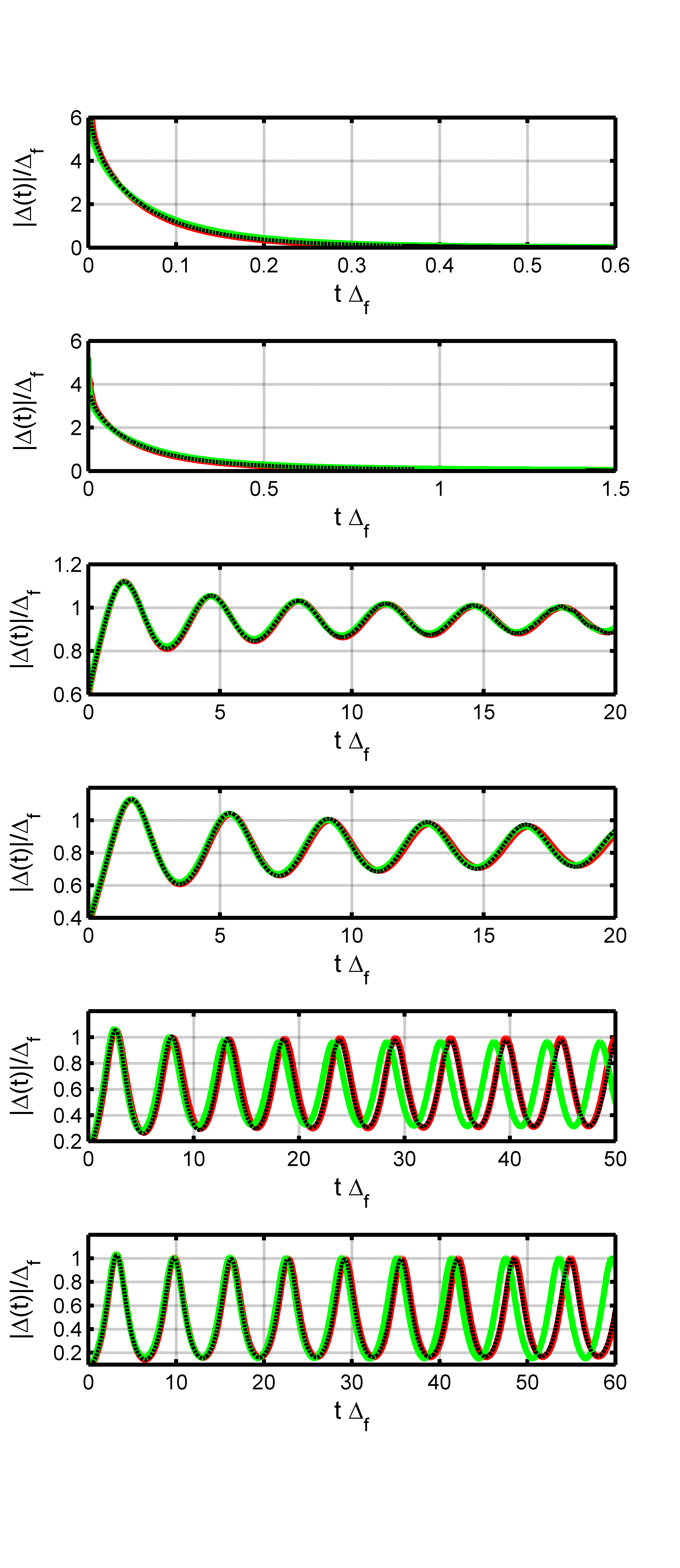}
	\caption{(Color online) Plot of gap dynamics for the non-integrable $d+id$ (full red line), the $s$-wave (dashed back line) and the exactly-integrable extended $d+id$ (full green line) for different quench parameters:  (a) ${\Delta_i}/{\Delta_f}=10$; (b) ${\Delta_i}/{\Delta_f}=5.2$, 
		(c) ${\Delta_i}/{\Delta_f}=0.5$, (d) ${\Delta_i}/{\Delta_f}=0.25$, (e) ${\Delta_i}/{\Delta_f}=0.05$ and (f) ${\Delta_i}/{\Delta_f}=0.02$. Calculations were performed on a system with $N=5024$ single particle energy levels and implemented an ultraviolet cutoff
		$\Lambda=4\eps_F$ for $d+id$-wave and $\Lambda=25\eps_F$ for s-wave dynamics.}
	\label{fig:didq}	
\end{figure}

In this section, we present the numerical results, obtained from the equations of motion (\ref{EqMot}), following a sudden change of pairing coupling $G$. To drive the system out-of-equilibrium, we take at $t=0$ the ground state of Hamiltonian (\ref{ASH}) with coupling constant $G_i$ and instantaneously change it to a different value of interaction $G_i \rightarrow G_f$, calculating the time evolution of order parameter from Eq. (\ref{EqMot}). As in the earlier studies \cite{Big-Quench-Review2015}, it is convenient to describe asymptotic states of the order parameter in terms of $\Delta_i$ and $\Delta_f$ - equilibrium order parameter values for $G_i$ and $G_f$ correspondingly.\\
In Fig. \ref{fig:didpd} we present quench phase diagram of time-dependent order parameter $\Delta(t)$ for the non-integrable $d+id$ ($G'(t)=0$) case. We have found out that the asymptotic states of $\Delta(t)$ at long times can be classified in terms of three dynamic regimes in the $(\Delta_i,\Delta_f)$ plane. For large $\Delta_i/\Delta_f$, we obtain regime (Region I) where the order parameter vanishes at long times (overdamped regime), Region II gives non-vanishing asymptote $\Delta(t\to\infty)=\Delta_\infty$ for $\Delta(t)$ and finally for sufficiently small $\Delta_i/\Delta_f$ we recover undamped oscillating phase in which $\Delta(t)$ oscillates between two limiting values. In passing, we note that the Region III is absent in recent calculation of the non-adiabatic pairing for $d$-wave superconductors ( $\Delta_\bk=\cos2\theta_\bk$) \cite{Capone2015} which is in stark contrast with our $d+id$-wave order parameter symmetry ($\Delta_\bk=\eps_k \Delta$). Presence of the nodal lines in the $d$-wave pairing leads to absence of the Region III \cite{Capone2015}.\\
In Fig. \ref{fig:didq}, we plot $|\Delta(t)|$ for different values of $\Delta_i/\Delta_f$ in weak coupling limit involving the $s$-wave, the non-integrable $d+id$ and the extended $d+id$ symmetry. Unlike the non-integrable case where quench of coupling affects (x,y) components of the field in Eq. \ref{Hmf}, all three components participate in a quench for the extended $d+id$ case through $\rho(t)$. As seen from Fig. \ref{fig:didq} that the extended $d+id$ dynamics model gives similar dynamics to the non-integrable $d+id$ case in terms of long time asymptotic states of $|\Delta(t)|$. Somewhat more surprising fact is that $|\Delta(t)|$ in the $s$-wave case also vary on similar time scales in weak coupling regime.  For values where $\Delta_i/\Delta_f\geq 5$ we get exponential damping of $\Delta(t)$ for the both $d+id$ cases again in full analogy with the $s$-wave and the $p+ip$ order parameters. For $\Delta_i/\Delta_f\ll 1$ we obtain the non-vanishing oscillations of $|\Delta(t)|$ between two limiting values.

\section{integrable exteded-$d+id$ phase diagram-Lax construction}
It has been shown in the earlier work \onlinecite{Links2013}, that the mean-field ground state of the extended $d+id$ model coincides with the exact solution of the Bethe anstaz in the continuum limit. Given that the mean-field ground state pseudospins are exact, we can exploit integrability of the extended-$d+id$ model to obtain exact asymptotic phases at long times. In this section we will show that all the three dynamical phases obtained following a quench of coupling in Fig. \ref{fig:didq}  generated by the mean-field BCS like Hamiltonian, are also present in the exactly-integrable - Lax vector method. Lax vector \cite{Big-Quench-Review2015}\cite{Yuzbashyan2013}\cite{Dzero-2dso-2015} for extended $d+id$ model is given by
\beg\label{Lax}
\begin{split}
\vec{L}(u)=\sum_{\bk}\frac{\eps_{\bk}\vec s_\bk}{u-\eps_\bk}-\frac{\vec e_z}{u{G\nu_F}} 
, \\
\end{split}
\en
{where $u$ is an arbitrary complex parameter. In Appendix we have shown that the square of the Lax vector is conserved by evolution. The conservation of ${\vec L}^2(u)$ allows one to determine asymptotic states of the order parameter depending on initial conditions.\cite{Big-Quench-Review2015} In order to compute the quench phase diagram at long-times, one needs to analyze complex roots of the spectral polynomial in the thermodynamic limit (for the definition of the spectral polynomial see e.g. Ref. \onlinecite{Big-Quench-Review2015}). For our model, the equation for the complex roots reads:
\beg\label{LAX_R}
\frac{\beta}{(1+\rho_0)-\frac{\mu_0}{u}\pm i\Delta_{0}}+\sum_{\bk}\frac{u\eps_{\bk}^2}{2(u-\eps_\bk)\sqrt{\tilde{\xi}_\bk^2+|\eps_\bk^2\Delta_{0}|^2}}=0,
\en
where we introduced parameters $\beta={g_f^{-1}}-{g_i}^{-1}$, $g=G\nu_F$ for brevity and $\Delta_0$, $\mu_0$ and $\rho_0$ denotes the ground state values obtained for coupling $g_i$.\\
In order to determine the steady state phase diagram we adopt the strategy described in Ref. \onlinecite{Big-Quench-Review2015}. Setting $u$ units of $\varepsilon_F$ along with other energies $\eps_{\bk}, \mu ,E_{\bk} $, noting that coupling $G$ has units $4\pi/k_F^4$ and expressing momentum in the units of the Fermi momentum $q=k/k_F$ we have
\beg\label{Eq4Roots}
\begin{split}
\frac{\beta}{u(1+\rho_0)-\mu_0\pm i\Delta_{0}u}+\int\limits_0^\infty\frac{\eps_{q}^2q{dq}}{2(u-\eps_{q})E(q)}=0,
\end{split}	
\en 
where $E(q)=\sqrt{[(1+\rho_0)\eps_q-\mu_0]^2+\eps_q^2\Delta_{0}^2}$.

To determine the boundaries separating various steady states we assume that the imaginary part of $u$ is infinitesimally small
\beg\label{uid}
u=v\pm i\delta. 
\en
Taking the real and imaginary parts of the equation (\ref{Eq4Roots}) with account of (\ref{uid}), we obtain two equations 
\beg\label{ReImEqs}
\begin{split}
&\frac{4\beta}{\pi}\Delta_{0}=v E(v), \\
&\frac{\pi v [v(1+\rho_{0})-\mu]\textrm{sign}\beta}{4E(v)\Delta_{0}}+\dashint\limits_{0}^\infty\frac{\eps_{q}^2qdq}{2(v-\eps_{q})E(q)}=0
\end{split}
\en
The results for the solution of these equations are presented in Fig. \ref{fig:extph}, where we show the asymptotic states phase diagram of the extended $d+id$ model obtained by the integrable method (Lax method). 
\begin{figure}[!h]
\includegraphics[width=8cm]{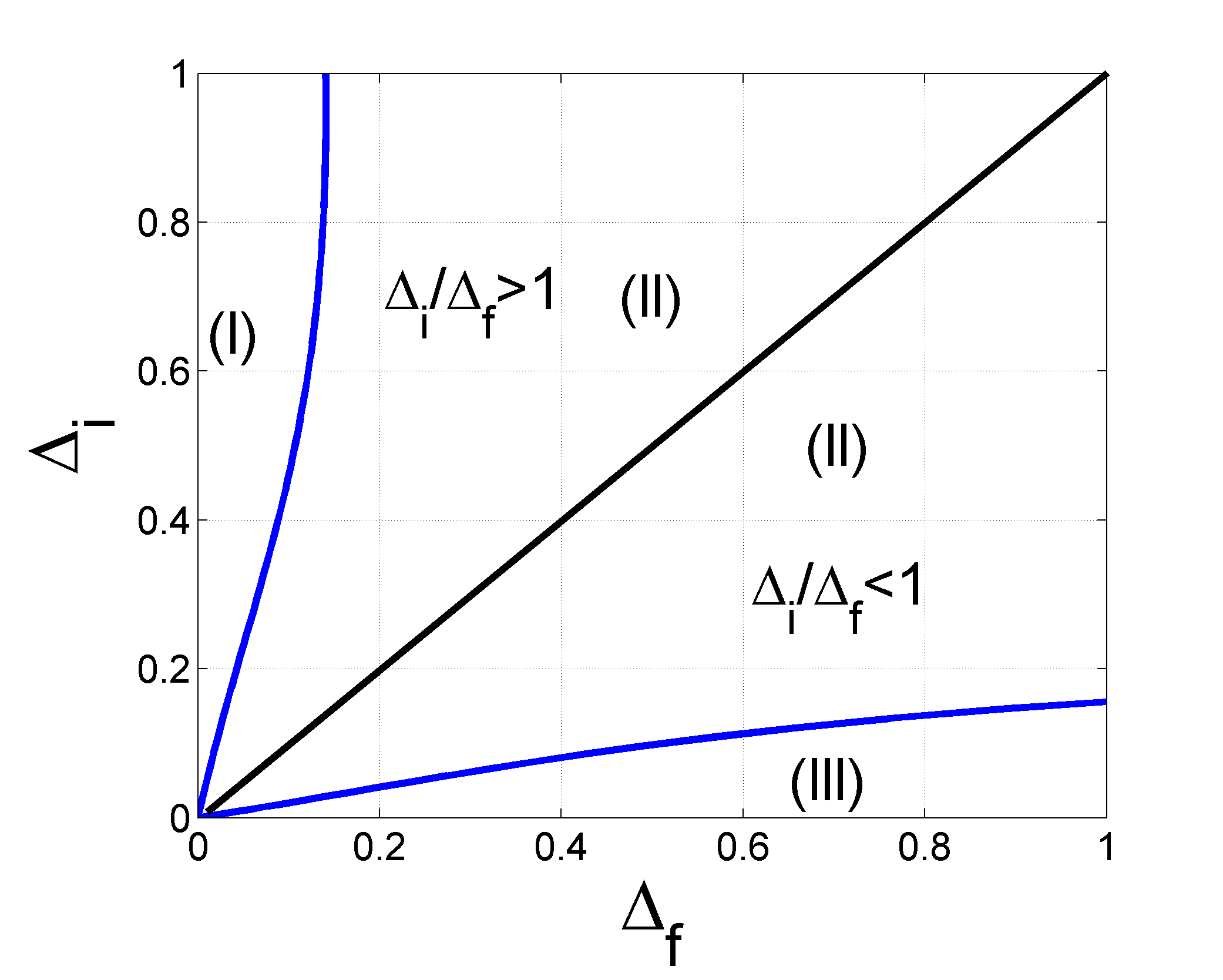}
\caption{Extended $d+id$ ($G'=G(t)$) exact phase diagram showing presence of all three dynamic regions. Computations have been performed for $\Lambda_E=4\eps_F$.}
\label{fig:extph}
\end{figure}
Comparing this diagram with the one found for the non-integrable $d+id$ model (Fig. \ref{fig:didpd}), it is clear that these two diagrams are quite similar to each other :
(i) for quenches corresponding to $\Delta_f\ll\Delta_i$ we find a gapless steady state in which the pairing amplitude vanishes;  (ii) for quenches when $\Delta_f\sim\Delta_i$ the pairing amplitude asymptotes to a constant and (iii) for quenches such that $\Delta_f\gg\Delta_i$ the pairing amplitude oscillated periodically and its time dependence is described by the Jacobi elliptic function. The Lax-mechanism gives the asymptotic dynamical phases from the mean-field ground state which is exact in thermodynamical limit \cite{Links2013} and the change of coupling, confirms presence of all three dynamical phases. This implies that the mean-field dynamics of the extended $d+id$ model obtained by computations are perfectly controlled and the fluctuations must cancel exactly at all levels. One can simply conclude same about the non-integrable $d+id$ case as limit $\rho\to0$ of the integrable extended-$d+id$ model.

\section{Discussion}
Immediately after the discovery of the special class of solutions which describe the pairing amplitude periodically oscillating in time provided the system is in the collisionless regime,\cite{Barankov2004} it was realized that the dynamics of the pairing amplitude as well as underlying pseudospin variables can be determined exactly.\cite{Emil2005a,Emil2005b,Emil2006} Naturally, the question of whether the steady states remain stable with respect to the integrability breaking perturbations were raised. The subsequent works, however, demonstrated the main features of the steady state diagram obtained from the exact integrability are retained (see e.g. Refs. [\onlinecite{BarankovGij,HanPu2014,Dzero-2dso-2015}]).
For example, the studies of quenched dynamics of two-dimensional spin-orbit coupled superfluids have shown that even for the quenches of the external Zeeman field lead to the asymptotic states found for the integrable $s$-wave pairing, including a state with the periodically oscillating pairing amplitude (in the latter case multiperiodic solutions may also appear).\cite{HanPu2014,Dzero-2dso-2015} The results presented here seem to confirm following general property: for the non-adiabatic dynamics integrability breaking perturbations have little effect on the resulting long-time dynamics phase diagram for the pairing with nodeless pairing amplitude in a sense that no qualitatively new steady states appear (or disappear) at long times. These results are applicable for zero temperature($T=0$) and when system has low energy excitations. Perhaps the most notable exception to this rule happens when the size of the system far exceeds the coherence length: in this case the steady state with the periodically oscillating $\Delta(t)$ develops spatial inhomogeneities driven by the parametric instability.\cite{Dzero2008}

While our results are perfectly applicable to systems consisting of charge neutral superfluids, at the level of the random-phase approximation, it can be demonstrated that the mean-field equations of motions found for the  problem without Coulomb interactions retain their form.\cite{Anderson1958} This statement is in agreement with a more qualitative argument based on the fact that the single particle relaxation time $\tau_\varepsilon$ far exceeds the characteristic time scale on which the order parameter evolves $\tau_\Delta$, so on the time scales $\tau_\Delta\ll t\ll \tau_\varepsilon$ the our pairing model with the reduced Hamiltonian should be valid. It has also been reported, in the context of high temperature cuprate ARPES experiments, that out-of-equilibrium quasiparticle populations exists even after the time scale of 5 p.s. (pico-seconds) \cite{Capone2015,RefE,RefF} whereas depending upon value of order parameter and fermi energy, the time scales involved in Fig.\ref*{fig:didq} are less than 1 p.s. Advent of femto-second probes will lead us to new horizons where not only theory of quantum quenches will be tested but rich information about pairing symmetries will be revealed

\section{Conclusions}
In this paper we have presented the results of our studies of out-of-equilibrium pairing dynamics in the $d+id$-wave and the extended $d+id$-wave models. We compared the resulting long time asymptotics for both of these models in which dynamics was initiated by a sudden change of the pairing strength and the initial state was always chosen to be system's ground state. We found that both phase diagrams turned out to be very similar despite the fact that the chiral $d+id$-wave model is not exactly integrable while the extended $d+id$ model is. Our work provides yet another example of a phenomenon for which insights obtained from exactly solvable models can be applied to describe the non-adiabatic dynamics of the pairing amplitude found for their non-integrable counterparts. We emphasize that s-wave like BCS dynamics in systems with BTRS and non-trivial ground state properties e.g. quantized boundary current of $I_B=2e\Delta/h$ and spontaneous magnetization etc., signal presence of the chiral $d+id$ pairing. By experimentally observing oscillating $\Delta(t)$ phase (phase III) in materials with otherwise d-wave pairing, will confirm mixing to $d+id$ pairing along with other unusual characteristics peculiar to the $d+id$ chiral superconductors. Thus, further pump probe experiments are needed to answer remaining questions of pairing symmetries.\\

\paragraph{Acknowledgments.} The authors are grateful to Emil Yuzbashyan for his comments on the manuscript and numerous stimulating discussions. We acknowledge the financial support by the National Science Foundation grant NSF-DMR-1506547.
The work of one of us (M.D.) was financially supported in part by the U.S. Department of Energy, Office of Science, Office of Basic Energy Sciences under Award No. DE-SC0016481. 

\begin{appendix}

\section{Integrals of motion for the extended $d+id$ model.}
In order to derive the integrals of motion, we use the method of Lax construction. 
The components of the Lax vector ${\vec L}(u)$ ($u$ is a parameter) are given by Eqs. (\ref{Lax}) in the main text. 
These quantities satisfy the algebra
\beg\label{LaxPoisson}
\begin{split}
\{L^+(u),L^-(v)\}&=-2\left[\frac{uL^z(u)-vL^z(v)}{u-v}\right], \\
\{L^z(u),L^+(v)\}&=-\left[\frac{uL^{+}(u)-vL^{+}(v)}{u-v}\right], \\
\{L^z(u),L^-(v)\}&=-\left[\frac{uL^{-}(u)-vL^{-}(v)}{u-v}\right].
\end{split}
\en
Note, that all three commutation relations retain the same form as in the $s$- and chiral $p$-wave cases.

Our main task now is to define the ``Casimir" of the Lax vector, $L_2(u)$, which will be conserved by the evolution. The dynamics of the Lax
vector components is described by the following equations which can be obtained from the equations of motion for the pseudospins together with Eq. (\ref{Lax}). 
For the dynamics of $\vec{L}(u)$ we find
\beg\label{Lplus}
\begin{split}
\frac{d{\vec{L}}}{dt}&= \textrm{det} \begin{bmatrix}
	\hat{x} & \hat{y} & \hat{z}  \\
	-2u\Delta_x & -2u\Delta_y & u\left(1-\frac{G}{\nu_F^{-1}}\sum\limits_{\bp}\varepsilon_\bp S_\bp^z\right)  \\
	L^x(u) & L^y(u) & L^z(u) 
\end{bmatrix}
\end{split}
\en
Where $\vec{L}\equiv\hat{x}L^x+\hat{y}L^y+\hat{z}L^z$ and $L^{\pm}=L^x\pm iL^y$. It is easy to see that quantity - the Lax norm - 
\beg\label{L2}
{L_2(u)=  L^+(u)L^{-}(u)+\left[L^z(u)\right]^2   }
\en
is conserved by the evolution i.e:
\beg
{\frac{dL_2(u)}{dt}=0.}
\en
In addition, the Poisson bracket which involves $L_2(u)$ is
{\small{\beg\label{L2uL2v}
\begin{split}
&\{L_2(u),L_2(v)\}=0.
\end{split}
\en}}
We will use this relation to show that the Hamilonian (\ref{Eq1}) is exactly integrable.

To show that number of the integrals of motion equals exactly to the number of the degrees of freedom, let us introduce the discreet mesh of momenta
\beg
\varepsilon_j=k_j^2/2, \quad (j=1,...,N)
\en
so that summation over the discreet energy levels $\varepsilon_j$ in the continuum limit become
\beg
\sum\limits_{j=1}^Nf(\varepsilon_j)\to\nu_F\int\limits_0^{{k_\Lambda^2}/{2}}f(\varepsilon) d\varepsilon,
\en
where $\nu_F=\frac{\cal A}{8\pi}$ is the two-dimensional single-particle density of states at the Fermi level. Hamiltonian in Eq.(\ref{ASHE}) can now be written as a spin chain

\beg\label{Hdis}
\begin{split}
	H&=\sum\limits_j\varepsilon_j2s_j^z-g\sum\limits_j\varepsilon_js_j^+\sum\limits_l\varepsilon_ls_l^- -g\sum\limits_j\varepsilon_js_j^z\sum\limits_l\varepsilon_ls_l^z
\end{split}
\en
and $g=G\nu_F$.
With these conventions the pseudospins are normalized:
\beg
\left({\vec s}_j\right)^2=\frac{1}{4}.
\en
For the Lax norm (\ref{L2}) we find
\begin{widetext}
\beg\label{LaxNorm}
\begin{split}
L_2(u)&=\sum\limits_{j=1}^{N}\sum\limits_{l=1}^{N}\frac{\varepsilon_j\varepsilon_l\left(s_j^{+}s_l^{-}+s_j^zs_l^z\right)}{(\varepsilon_j-u)(\varepsilon_l-u)}+\frac{2}{ug}\sum\limits_{j=1}^N\frac{\varepsilon_js_j^z}{\varepsilon_j-u}+\frac{1}{u^2g^2}=\sum\limits_{j=1}^N\frac{H_j}{u-\varepsilon_j}+\frac{1}{4}\sum\limits_{j=1}^N\frac{\varepsilon_j^2}{(u-\varepsilon_j)^2}+\frac{J_z}{gu}+\frac{1}{u^2g^2},
\end{split}
\en
\end{widetext}
where $H_j$ denotes the Hamiltonian
\beg\label{Hj}
H_j=-\frac{2s_j^z}{g}+\sum\limits_{l\not=j}\frac{\varepsilon_j\varepsilon_l}{\varepsilon_j-\varepsilon_l}\left(s_j^+s_l^-+s_j^-s_l^++2s_j^zs_l^z\right)
\en
and $J_z$ gives the total pseudospin projection on $z$-axis
\beg\label{Jz}
J_z=2\sum\limits_{j=1}^Ns_j^z=\textrm{const}.
\en
Since $L_2(u)$ is conserved by the evolution so that $\{L_2(u),H_j\}=0$, equation (\ref{L2uL2v}) implies that $\{H_i,H_j\}=0$, i.e. $H_i$'s are mutually conserved. There are $N$ independent $H_j$'s in a system of $N$ spins. Therefore, we have identified $N$ integrals of motion for a system of $N$ spins. Furthermore, our initial Hamiltonian (\ref{Hdis}) can be expressed in terms of $H_j$'s as follows
\beg\label{HdisHj}
{H=-g\sum\limits_{j=1}^N \eps_j H_j}+\textrm{const.}
\en
This equations means that ${H}$ Poisson commutes with $L_2(u)$ and we have identified all integrals of motion. Hence, the extended $d+id$ model is exactly integrable.
\end{appendix}

\bibliography{didbiblio1}

\end{document}